\documentclass[conference,10pt]{IEEEtran}
\usepackage{epsfig,epsf,rotating,setspace,latexsym,amsmath,amssymb,amsfonts,bm,theorem,subfigure,epstopdf,cite,authblk,bbm,nonfloat,comment}
\usepackage{algorithm}
\usepackage[noend]{algpseudocode}
\usepackage{color}
\usepackage{mathtools}
\usepackage{soul}

\algrenewcommand\algorithmicforall{\textbf{foreach}}
\algrenewcommand\algorithmicindent{.8em}

\newenvironment{Proof}[1]{\medskip\par\noindent{\bf Proof:\,}\,#1}{{\mbox{\,$\blacksquare$}\par}}

\newtheorem{defn}{Definition}
\newtheorem{thm}{Theorem}
\newtheorem{lem}{Lemma}

\newtheorem{obs}{Observation}

\newcommand{\eps}{\varepsilon}
\renewcommand{\epsilon}{\eps}
\renewcommand{\P}{\mathbb{P}}

\newcommand{\E}{\mathbb{E}}

\newcommand{\cT}{\mathcal{T}}
\newcommand{\cF}{\mathcal{F}}

\newcommand{\cN}{\mathcal{N}}

\IEEEoverridecommandlockouts
\allowdisplaybreaks
\begin{document}

\title{Information Degradation and Misinformation \\ in Gossip Networks} 

\author{Thomas Jacob~Maranzatto \qquad Arunabh Srivastava \qquad Sennur Ulukus\\
        \normalsize Department of Electrical and Computer Engineering\\
        \normalsize University of Maryland, College Park, MD 20742\\
        \normalsize  \emph{tmaran@umd.edu} \qquad \emph{arunabh@umd.edu} \qquad \emph{ulukus@umd.edu}}

\maketitle

\begin{abstract}
    We study networks of gossiping users where a source observing a process sends updates to an underlying graph.  Nodes in the graph update their neighbors randomly and nodes always accept packets that have newer information, thus attempting to minimize their age of information (AoI).  We show that while gossiping reduces AoI, information can rapidly degrade in such a network.  We model degradation by arbitrary discrete-time Markov chains on $k$ states.  As a packet is transmitted through the network it modifies its state according to the Markov chain. In the last section, we specialize the Markov chain to represent misinformation spread, and show that the rate of misinformation spread is proportional to the age of information in both the fully-connected graph and ring graph.
\end{abstract}

\section{Introduction}
Given a network of $n$ unreliable users, can we characterize the accuracy of information stored at a single user?  This paper aims to study this question when the users communicate using a gossiping scheme. Here, a source $n_0$ generates and sends updates to a network $G = (\cN, E)$, and nodes/users in the network randomly share their status with their neighbors.  Packets are timestamped, and when a newer packet reaches some user, the user accepts the new packet and discards their old packet. When the source sends an update to some user, that information has the highest possible quality. Furthermore, as a packet is disseminated in the network, its quality may degrade according to a separate process that depends only on the number of edges the packet has traversed. We formalize our model in Section~\ref{sec:model}.

This simple protocol of randomly distributing a packet to a neighbor is known as \textit{gossiping}. Gossiping has been implemented in databases as a way to keep up-to-date records since at least the 1980s~\cite{database_gossip}, and has proven useful in other domains~\cite{gossip_promise}. Recently, a line of work has emerged investigating the \textit{age of information} (AoI) in gossip networks. Roughly, age of information is a measure in how out-of-date the information at a single user/node is compared to some global source. In two works of Yates~\cite{yates_gossip_isit, yates_gossip_spawc}, he showed that the AoI of the fully-connected network scales logarithmically with the number of users. Further work by Buyukates, Bastopcu, and Ulukus~\cite{Buyukates_2022} showed that the AoI of a ring network scales as $\sqrt n$, and Srivastava and Ulukus~\cite{gossip_grid} showed that the AoI on a square lattice (grid) scales as $n^{1/3}$. Maranzatto~\cite{age_bipartite} also studied random and bipartite networks, and proved the existence of a phase transition for AoI in $G(n,p)$. Maranzatto and Michelen~\cite{age_percolation} showed that the AoI of a user is equivalent to a \emph{first-passage percolation} process assuming the modeling assumptions made originally by Yates, and gave tight upper and lower bounds for the expected AoI in arbitrary networks.

\begin{figure}
    \centering
    \includegraphics[width=\linewidth]{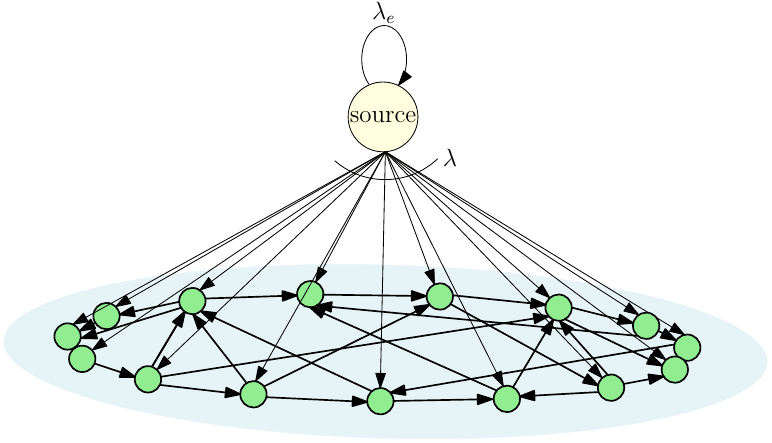}
    \caption{A gossiping network with one source node generating and sharing updates. The nodes in the network share information with each other.}
    \label{fig: gossip only}
\end{figure}

This line of work has shown that, under the modeling assumptions discussed in Section~\ref{sec:model}, networks with high connectivity are able to maintain sub-linear, and in some cases logarithmic, AoI scaling with the number of users. This leaves open questions in information quality, as in large networks, a packet stored at a user has lasted on average a non-constant amount of time. In this interval of time the packet may have incurred some corruption, either at random or by a malicious attack. In previous work, Kaswan and Ulukus~\cite{Kaswan_mutation} studied numerically the tradeoff between the age of information (AoI) and average proportion of nodes containing misinformation. Their work uses the tools developed by Yates~\cite{yates_gossip_isit} which involves unrolling certain recursive functions on the graph.  The recursive functions are tractable in the AoI setting, but become unwieldy when tracking misinformation.  We take a different approach, using first-passage percolation as our main tool to study the degradation of information.  The proof of our main result in Section~\ref{sec:general_proof} follows a similar proof technique found in~\cite{age_percolation}.

Our contribution is twofold.  First, in Section~\ref{sec:general_proof} we generalize from misinformation to arbitrary Markovian models of information degradation, thus allowing for more fine-grained analysis of the quality of information stored by any user.  Second, in Section~\ref{sec:misinfo} we specialize the degradation process to have two states, ``True'' and ``False'', and give explicit bounds on misinformation spread in two networks.  We study the fully connected network $K_n$, and the ring network $C_n$, and show that the proportion of users with true information is much higher in $K_n$ compared to $C_n$.  In fact, the rate of information degradation is exponential in the age of information in both networks, and we conjecture that this is true for all gossip networks.  If this conjecture holds, then the networks with the best AoI will spread misinformation the slowest.

\begin{figure}
    \centering
    \includegraphics[width=\linewidth]{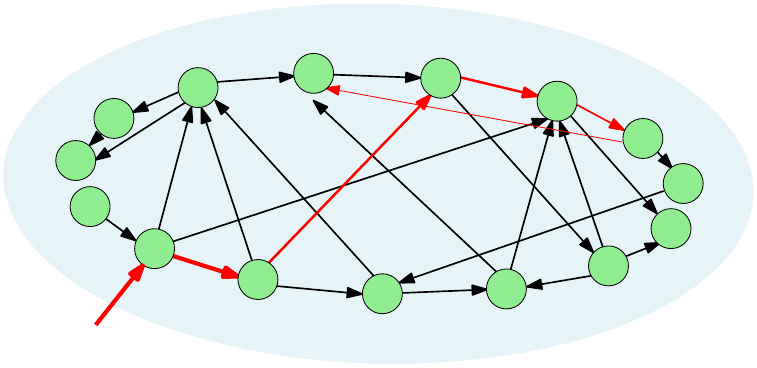}
    \caption{This figure shows the process of information degradation with time, which we follow with the transfer of a packet to nodes in the network starting from the source. The thinning of the red arrow represents the degradation of the packet being shared between nodes. In this gossiping network, the source sends an update to a node, represented by the red arrow coming from outside. This information is not degraded since it comes directly from the source, thus being represented by the thickest arrow. When this node shares information with the next node, there is information degradation, depending on the associated Markov chain.}
    \label{fig: degradation figure}
\end{figure}

\section{System Model and Preliminaries}\label{sec:model}
\subsection{Age and Quality of Information}
We begin by describing the communication model, where nodes in a graph send natural numbers which can for instance represent versions of a piece of software.  Consider a source node $n_0$ which sends updates to a network $G = (\cN, E)$ with $n$ nodes.  We let $\cN = [1,\ldots, n]$.  The source receives version updates via a Poisson process with rate $\lambda_e$, and sends updates to each $v \in G$ as independent Poisson processes with identical rates $\frac{\lambda}{n}$.  In all graphs we consider, an undirected edge $(i,j) \in E$ enables two-way communication between nodes $i$ and $j$ following two independent Poisson processes, with $\lambda_i(j) = \frac{\lambda}{\operatorname{deg}(i)}$ denoting the Poisson rate from $i$ to $j$, and $\lambda_j(i) = \frac{\lambda}{\operatorname{deg}(j)}$ the rate from $j$ to $i$, where $\operatorname{deg}(i)$ denotes the degree of node $i$.  Note that in general $\lambda_j(i) \not= \lambda_j(i)$.  All graphs we consider are connected, thus we will never have the edge case $\operatorname{deg}(i) = 0$.   The source and every node in the network contain internal counters; when a node $i \in  \cN \cup \{n_0\}$ communicates to a neighbor $j$ (because $i$'s Poisson process updated) $i$ sends its current counter value.  The counter for $n_0$ increments if and only if the process for $n_0$ updates.  Contrast this with $j \in \cN$ whose counter increments if and only if $j$ receives a newer version from one of its neighbors.  Let $X_j(t)$ be the number of versions node $j$ is behind $n_0$ at time $t$.

In addition to the internal version age counters, each node also contains a variable $i \in [0,\ldots, k-1]$ representing the quality of information it stores.  This could for instance model the number of errors in a file, or as we see in Section~\ref{sec:misinfo} an indicator that a file contains misinformation.  These variables represent states in a discrete time Markov chain $M = (\mathcal{S}, T)$ with states $\mathcal{S} = [0,\ldots, k-1]$ and transitions $T$.  Then, at all times, node $i$ stores (a copy of)  some state in $\mathcal{S}$.  Node $n_0$ always stores the initial state, which by convention we take to be $0$.  Let $S_i(t)$ denote the state stored by node $i$ at time $t$; we call $S_i(t)$ the \textit{quality of information}. Let $M_s(r)$ denote the distribution on $\mathcal{S}$ induced by running $M$ for $r$ steps starting in state $s$.  Then, when node $i$ receives an update from node $j$ at time $t$, one of two outcomes is realized:
\begin{itemize}
    \item If $X_i(t^-) < X_j(t^-)$, then $S_i(t) = M_{S_j(t^-)}(1)$.
    \item Else $S_i(t) = S_i(t^-)$.
\end{itemize}
In words, if $i$ receives a newer packet, it immediately accepts it, but the quality of the received packet changes according to the chain $M$.  If $i$ receives a packet with the same or worse age, it rejects it and maintains its current information.  If a node has not received any packet at time $t$, then we choose to let $S_i(t) = -1$; this choice is arbitrary and will not impact our results.

\subsection{Percolation and Hopcount}
First-passage percolation is a model for fluid flow through a random porous medium.  There is an expansive literature on first-passage percolation in different contexts; we direct the reader to~\cite{auffinger2017fifty} for an excellent survey and compilation of open problems.  For our purposes, we only need to consider finite graphs with independent random edge weights.  Let $G = (V,E,W)$ be a finite directed weighted graph.  The weights $W = \{\tau_e \}_{e \in E}$ are chosen according to independent probability distributions.  Let $\gamma = (e_1,\ldots,e_m)$ be some path in $G$.  The passage time of $\gamma$ is
\begin{align}
T(\gamma) := \sum_{e \in \gamma}\tau_e.
\end{align}
Then, for any two nodes $i,j \in \cN$, the \emph{passage time} from $i$ to $j$ is
\begin{align}
T(i,j) := \min_\gamma T(\gamma),
\end{align}
where the minimum is over paths $\gamma$ which start at $i$ and end at $j$. Finally, define the \emph{hopcount} from $i$ to $j$ as the number of edges in the path realizing the minimum in $T(i,j)$, and denote this by $H(i,j)$.

Let $G$ be a gossip network with source $n_0$.  We define the weighted \emph{auxiliary graph} $G' = (\cN \cup n_0, E', W')$ on the same vertex set as $G$ plus the source with the following edges and weights:
\begin{itemize}
    \item The edge set $E' = \{(i,j) : (j,i) \in E \}$, thus all edges are reversed in $G'$.
    \item The weights $W'= \{\tau_{(i,j)}\}_{(i,j) \in E'}$ are independent exponential random variables with parameter $\frac{\lambda}{\operatorname{deg}(j)}$.
\end{itemize}

Thus, the graph $G'$ is just $G$ with all edges reversed, and the Poisson process on edges are replaced with exponential random variables.

Then, following the work of~\cite{age_percolation}, our first result is that the quality of information is a function of the hopcount in $G'$ and the mixing of the Markov chain $M$.  If a node has not received a packet at time $t$, we set $S_i(t) = -1$.

\begin{thm}\label{thm:general}
    For every time $t$ and node $i$, the quality of information of $i$ at time $t$ has distribution given by
    \begin{align}
        S_i(t) = \begin{cases}
        -1, & \text{ if } \ T(i, n_0) > t, \\
        M_0(H(i, n_0)), & \text{ else}.
    \end{cases}
    \end{align}
\end{thm}

In words, the quality of information at node $i$ is undefined (set to -1) if a packet has not yet reached $i$ at time $t$, otherwise the quality of information is obtained by running $M$ for the number of edges in the shortest source-to-$i$ path.  We prove Theorem~\ref{thm:general} in Section~\ref{sec:general_proof}

In Section~\ref{sec:misinfo}, we specialize the quality of information Markov chain to two states $\mathcal{S} = \{\cT, \cF\}$, representing true and false information, respectively.  The motivation for this is to study a simplified model of information mutation, where each node forwards truthful information with probability $p$, and otherwise ``lies'' and forwards false information.  This is similar to the model considered by Kaswan and Ulukus~\cite{Kaswan_mutation}, however, in their model nodes will accept incoming \textit{truthful} packets with the same version age.  In their model ``truth prevails'', and it is not hard to show that the proportion of nodes containing true information in their model stochastically dominates our model.\footnote{Let $G$ be a gossip network, and consider the natural coupling between the two models.  Then, if node $i$ contains true information at time $t$ in our model, this implies $i$ contains true information in the model proposed in~\cite{Kaswan_mutation}.}  Our model is more pessimistic, and assumes that ``timeliness prevails'', in that nodes will always prefer the information that first reaches them.  Finally, we remark that the computations bounding the hopcount for $C_n$ in this section are inspired by similar techniques found in~\cite{gossip_grid} and~\cite{Buyukates_2022}.

\section{Proof of Theorem~\ref{thm:general}}\label{sec:general_proof}
To prove Theorem~\ref{thm:general}, we follow similar techniques to the proof of~\cite[Theorem~2]{age_percolation}, which we restate now.  Recall that the \textit{age process} of a vertex $i \in \cN$ is defined as $\Tilde{X}_i(t) := t - N_i(t)$, where $N_i(t)$ is a real number indicating the exact time the packet stored at $i$ was generated by the source.  $\Tilde{X}$ is just the continuous analog of the version age of information.

\begin{thm}[\cite{age_percolation}]\label{thm:aoi}
    For any $i \in \cN$ and $t \geq 0$, the \textit{age process} of $i$ at time $t$ has distribution given as,
    \begin{align}
        \Tilde{X}_i(t) = \min\{T(i, n_0), t\}.
    \end{align}
\end{thm}

The main difference in our setting is considering hopcount instead of first-passage time, but the main idea is the same.  The discrete-time Markov chain $M$ is only invoked when a node receives a fresher packet, thus, since the shortest time path from $i$ to the source in $G'$ at time $t$ is equal in distribution to the age process, the packet transmitted on this path will be updated by the Markov chain exactly the number of hops on this path.  We formalize this intuition now.

\begin{Proof}[\textit{Theorem~\ref{thm:general}}]
We give an exact characterization of the minimizing path in terms of $G'$.  Consider a modified process where the Poisson process edges in $G$ are reversed, and a single packet is generated at some node $i \in \cN$.  In other words, $i$ will act as the source and generate exactly one packet at time $t=0$; let $X'_i(t)$ be the age process of $n_0$ in this new model. By definition, if $n_0$ received $i$'s packet at time $t'$, then at $t \geq t'$ the age is $X'_{i}(t) = t - t'$, otherwise it is $+\infty$.  We claim that $\inf_{t \geq 0}\{X'_i(t) \not= \infty\}$ is equal in distribution to the first passage time $T_{G'}(i,S)$.  To be explicit consider the following quantities, 
\begin{align}
    \partial_t &:= \Big\{(u,v) : X'_u(t) = 0,  X'_v(t) = +\infty\Big\}, \label{eqn:sphere}\\
    t^+ &:= \inf_{t' > t} \Big\{\text{some node with age $+\infty$ receives} \nonumber\\
    &\quad\quad\quad\quad \text{the packet from $i$ at time $t'$} \Big\}. \label{eqn:updatetime}
\end{align}
    
With probability 1, $t^+ \not = t$.  Let $w^{(u,v)}$ be the Poisson rate for edge $(u,v)$ in the modified process; observe this is equal to the parameter in the exponential distribution for edge $(u,v)$ in graph $G'$.  Furthermore, the event in \eqref{eqn:updatetime} involves exactly those edges in the boundary $\partial_t$ of the random ball $B_j(t)$, thus by the strong Markov property of the Poisson process,
\begin{align}
    t^+ = \inf_{(u,v) \in \partial_t} \left\{w^{(u,v)}\right\}.
\end{align}
If $\partial_t \not= \emptyset$, then $\partial_t \not= \partial_{t^+}$, thus the boundary evolves according to the inter-arrival distributions that correspond to the inverted edges in $G'$.  Therefore, in the modified process, the age of information at a node $n_0$ has distribution given by, 
\begin{align} \label{eqn:noupdate}
    X'_i(t) = \begin{cases}
        t - T_{G'}(i,S), &\text{ if } T_{G'}(i,S)\leq t, \\
        +\infty, &\text{ otherwise }.
    \end{cases}
\end{align}

Note that \eqref{eqn:noupdate} shows that $T_{G'}(i,S)$ is the amount of time that passes between $i$ sending a packet and $n_0$ receiving it.  Again, due to the strong Markov property of the Poisson process, this time is independent of when the packet is sent.  Now, in the original model a node only accepts a packet if it contains fresher data and rejects the packet otherwise, thus stale packets will not interact with new packets.  Now, let $\gamma'$ be the path in $G'$ that achieves $T_{G'}(i, S)$.  If no such path exists, then $X'_i(t) = +\infty$, thus we can set $S_i(t) = -1$. If the path exists, then with probability 1 $\gamma'$ is unique.  Note that $\gamma'$ corresponds to a path $\gamma \in G$ which is the same, but all edges are reversed; in particular both paths have the same length.  Then, any packet that is sent from $n_0$ to $i$ along $\gamma$ will have its quality of information mutated by the Markov chain $M$ exactly equal to the length of $\gamma'$, assuming the packet is not overwritten at some point.  But by Theorem~\ref{thm:aoi}, $T_{G'}$ is equal in distribution to the age process (for large enough $t$), thus the packet sent from $n_0$ to $i$ in $G$ along $\gamma$ can be coupled to the corresponding path $\gamma'$ and will survive.  This implies the surviving packet will be mutated by the Markov chain $M$ exactly equal to the number of edges in $\gamma$.  Since $\gamma$ is unique w.p. 1, this is just the hopcount, and hence $S_i(t) = M_0(H(i,n_0))$.  Combining this with the case that a packet has not reached $i$ from $n_0$ discussed above completes the proof.
\end{Proof}

\section{Misinformation: Comparison of $K_n$ and $C_n$}\label{sec:misinfo}

Inspired by the study of misinformation spread in networks~\cite{Kaswan_mutation}, we specialize the Markov chain $M$ to have two states $\cT$ and $\cF$ with the following transitions:

\begin{table}[h]
    \centering
    \begin{tabular}{l|cc}
    & $\cT$ & $\cF$ \\ \hline
    $\cT$ & $p$ & $1-p$ \\
    $\cF$ & $0$ & $1$                 
    \end{tabular}
\end{table}

Here, $p$ may vary with the number of vertices $n$, and the initial state is taken to be $\cT$. From now on, when referring to the Markov chain $M$ we mean the chain on states and transitions defined above.  We specialize to this chain in particular because it provides a nice model for studying misinformation spread on a network.  On each re-transmission of a truthful packet, there is a $1-p$ probability the packet is corrupted to be false.  Once a packet is false, it cannot recover to be truthful.  

Since nodes only accept fresher packets, we heuristically expect the distribution of nodes with true versus false information to behave as follows:  At large, fixed $t$ there exist some subset of vertices $U \subset \cN$ where any $u \in U$ holds a packet that was directly sent to it by the source.  By the definition of the model, these vertices hold truthful information.  Surrounding any of these $u$ is a random ball containing truthful nodes with the same version age.  This is contained in a larger ball where nodes hold false information but have the same version age.  Since new packets override old packets, these true/false balls surrounding the $u$ intersect and override each other.  Indeed, a vertex $v$ may have contained truthful information at time $t^-$, but at time $t$ accepted a newer packet with false information.  Eventually a packet sent far in the past will no longer be present in the network.

Recall that $M_{\cT}(k)$ is the distribution on $\{\cT, \cF\}$ by running $M$ for $k$ steps starting in state $\cT$.  We need the following easy observation.
\begin{obs}\label{obs:markov_mix}
    \[\P[M_{\cT}(k) = \cT] = p^k,\]
    and
    \[\P[M_{\cT}(k) = \cF] = 1 - p^k.\]
\end{obs}

For any time $t$, let $P_G(t) = \frac{1}{n}\sum_{i \in \cN}\mathbbm{1}\{S_i(t) = \cT\}$ be the proportion of nodes containing truthful information at time $t$ for gossip network $G$.  A key quantity of interest is the limiting average number of nodes that contain truthful information. For any graph $G$ define,
\begin{align}\label{eq:percentage}
    P(G) := \lim_{t \to \infty} \E\big{[}P_G(t)\big{]}.
\end{align}
Then, $P(G)$ gives a coarse view of the dynamics of $M$ on $G$.  In light of Theorem~\ref{thm:general}, to bound $P(G)$ we need to bound the hopcount in the auxiliary graph $G'$, and bound the mixing time of $M$.  Furthermore, the time limit ensures the system is at equilibrium. We could have equivalently taken the smallest time such that every node in the network has received a packet.

Let $K_n$ and $C_n$ be fully-connected and the grid gossip networks on $n$ vertices, respectively.  For notational ease, we set $\lambda = 1$, but the results are easily extended to arbitrary $\lambda$.  The authors in~\cite{BHAMIDI_2011} show that the expected hopcount between any two vertices in $K_n$ with i.i.d. $\operatorname{Exp}(1)$ edge weights is asymptotic to $\log n + \gamma - 1$, where $\gamma$ is Euler's constant.  In our model, the edge weights in the auxiliary graph depend on if one of the endpoints of $(i,j)$ is the source vertex.  If neither vertex is  the source, then the edge weight is distributed as $\operatorname{Exp}\left(\frac{1}{n-1} \right)$.  Otherwise, the edge weight is 0 if $i$ is the source, and $\operatorname{Exp}\left(\frac{1}{n} \right)$ if $j$ is the source.

Notice that any path achieving the first-passage time will never traverse an edge twice, thus the $0$ weight edges can be changed to $\operatorname{Exp}\left(\frac{1}{n} \right)$ weight edges while the hopcount from $i$ to $n_0$ remains the same.  Furthermore, while rescaling all edge weights by a factor $n$ will rescale the first-passage time, the hopcount will remain the same.  Therefore, for large enough $n$, the expected  hopcount for the gossip network $K_n$ is $\Theta( \log n)$.  We omit the tedious details, as the proof follows almost verbatim from~\cite{BHAMIDI_2011} and involves an analysis of a coupled branching process.  Their result, combined with Observation~\ref{obs:markov_mix} gives the following result.

\begin{thm}\label{thm:k_n}
    The average number of nodes containing truthful information in $K_n$ is given by,
    \begin{align}
        P(K_n) = p^{\Theta( \log n)}.
    \end{align}
\end{thm}

Therefore, for any $C \in (0,1)$, as long as $p = C^{- O( \log n)}$ then $P(G) \geq C$; a constant fraction of vertices hold true information.

We now focus on bounding the hopcount for the auxiliary cycle gossip network.
\begin{lem}\label{lem:cycle}
    For any $i \in [n]$, the expected hopcount for the auxiliary ring gossip network satisfies,
    \begin{align}
        \E H(i, n_0) = \Theta(\sqrt n ).
    \end{align}
\end{lem}

\begin{Proof}
      Let $C'_n$ be the auxiliary graph for the ring gossip network. The edges $(i, i+1)$ and $(i, i-1)$ (taken mod $n$) in $C'_n$ have rates $\operatorname{Exp}(1/2)$, and the edges $(i, n_0)$ have rate $\operatorname{Exp}(1/n)$.  Consider a discrete time Markov chain $\{X(t)\}_{t \in \mathbb{N}}$ on transient states $[0,\ldots, n-1]$ and a single absorbing state $n$.  The transition from state $i$ to state $j$ is given by,
      \begin{align}
      p_{ij} = \begin{cases}
        \frac{n-i}{n+1}, & \text{if } j = i+1 \text{ and } j \not= n,\\
        \frac{i+1}{n+1}, & \text{if } j = n \text{ and } i\not= j,\\
        0, & \text{ else.}
        \end{cases}
    \end{align}
    Notice that there is a coupling between the number of transitions before $\{X(t)\}$ absorbs and number of unique states that can be visited in $C'_n$ before the first-passage time $T(0, n_0)$.  The probability that the chain transitions from state $i$ to $i+1$ in $\{X(t)\}$ is exactly equal to the probability that a non-source state has a higher probability of being visited than the source state in $C'_n$, given that $i$ non-source states have already been visited.  Let $B(t)$ be the set of states in $C'_n$ that can be reached from state $0$ with total path weight less than $t$, and let $\partial(B(t))$ be the set of edges with one endpoint in $B(t)$.  Let $\{Y_i\}$ be a collection of  i.i.d. $\operatorname{Exp}(\frac{1}{2})$ random variables, and $\{Z_i\}$ a collection of i.i.d $\operatorname{Exp}(\frac{1}{n})$ random variables. Then,
    \begin{align}
        \P[&\arg\min_j\{(i,j) \in \partial(B(t))\} \not= n_0 : |B(t)| = i] \nonumber\\
        &= \P[\min_{i \in \{0,1\}}\{Y_i\} < \min_{j \in [i]}\{Z_j\}]\\
        &= \P[\operatorname{Exp}(1) < \operatorname{Exp}(\frac{i}{n})]\\
        &= \frac{n-i}{n+1}.
    \end{align}
    Therefore, the hopcount is bounded above by the number of transitions before $\{X(t)\}$ absorbs.  Now, we compute the expected number of transitions before absorption,
    \begin{align}
        \E[&\text{absorbing time of } \{X(t)\}] \nonumber\\
        &=\sum_{j=1}^n j \cdot \P\big[\{X(t)\} \text{ absorbs after exactly $j$ transitions}\big]\\
        &= \sum_{j=1}^n \frac{j^2}{n+1}\prod_{i=0}^{j-2}\frac{n-i}{n+1}\\
        &\approx \sum_{j=1}^n \frac{j^2}{n}\prod_{i=0}^{j-2}\left(1 - \frac{i}{n}\right)\\
        &= \sum_{j=1}^n \frac{j^2}{n}\exp\left( \log \left(\prod_{i=0}^{j-2}\left(1 - \frac{i}{n}\right) \right) \right)\\
        &\leq \sum_{j=1}^n \frac{j^2}{n}\exp\left( -\sum_{i=0}^{j-2}\frac{i}{n} \right)\\
        &\approx \sum_{j=1}^n \frac{j^2}{n}\exp\left( -\frac{j^2}{n} \right),
    \end{align}
    where the inequality follows from $\log(1+x) \leq x$ when $x > -1$, the first approximation is by replacing $n+1$ with $n$, and the last approximation is ignoring lower order terms.  Now, using a Riemann sum approximation with step size $\frac{1}{\sqrt{n}}$ and letting $n \to \infty$,
    \begin{align}
        \frac{1}{\sqrt n} \sum_{j=1}^n \frac{j^2}{n}\exp\left( -\frac{j^2}{n} \right) &= \int_0^\infty x^2 e^{-x^2} dx\\
        &= \frac{\sqrt \pi}{4},
    \end{align}
    which implies $\E H(i, n_0) \leq \frac{\sqrt {n \pi}}{4}$. An analogous argument for the directed ring with edge weights as $\operatorname{Exp}(\frac{1}{2})$ gives $\E H(i, n_0) \leq \frac{\sqrt {n \pi}}{8}$, which is sufficient.
\end{Proof}

An immediate corollary of Lemma~\ref{lem:cycle} and Observation~\ref{obs:markov_mix} is the following characterization of the proportion of truthful nodes.

\begin{thm}
    The average number of nodes containing truthful information in $C_n$ is given by,
    \begin{align}
        P(C_n) = p^{\Theta(\sqrt n)}.
    \end{align}
\end{thm}

Therefore, for any $C \in (0,1)$, if $p = C^{-O(\sqrt n)}$ then $P(G) > C$; a constant proportion of nodes contain truthful information in expectation.

\section{Remarks}
While in both the fully-connected and ring networks the proportion of truthful nodes decreases with $n$, the rate of decrease is exponentially slower for the fully-connected network. In both networks, the hopcount scales at the same rate as the first-passage time, hence this is expected. We leave as an interesting open problem the question of showing that all gossip networks have hopcount and first-passage time scaling at the same rate.  

We also note that our results match the predictions made in simulations performed by~\cite{Kaswan_mutation}, where they show that decreasing $p$ increases the proportion of vertices holding misinformation.  However, in their simulations, even for large $p$ some positive fraction of the vertices hold truthful information.  We believe this is because in their model, stale packets with the same version age can interact and convert false packets to true packets as ``truth prevails'' there.  Therefore, as long as the rate from the source is slow enough, and the network facilitates information flow, some proportion of the vertices will contain truthful information.  It would be interesting to study their more optimistic model in the first-passage framework, but the interactions of packets on the boundary of the first-passage ball make analysis more delicate.

\bibliographystyle{unsrt}
\bibliography{bibliography}

\begin{thebibliography}{10}

\bibitem{database_gossip}
A.~Demers, D.~Greene, C.~Hauser, W.~Irish, J.~Larson, S.~Shenker, H.~Sturgis, D.~Swinehart, and D.~Terry.
\newblock Epidemic algorithms for replicated database maintenance.
\newblock In {\em ACM PODC}, page 1–12, 1987.

\bibitem{gossip_promise}
K.~Birman.
\newblock The promise, and limitations, of gossip protocols.
\newblock {\em ACM SIGOPS Oper. Syst. Rev.}, 41(5):8–13, October 2007.

\bibitem{yates_gossip_isit}
R.~D. Yates.
\newblock The age of gossip in networks.
\newblock In {\em IEEE ISIT}, July 2021.

\bibitem{yates_gossip_spawc}
R.~D. Yates.
\newblock Timely gossip.
\newblock In {\em IEEE SPAWC}, 2021.

\bibitem{Buyukates_2022}
B.~Buyukates, M.~Bastopcu, and S.~Ulukus.
\newblock Version age of information in clustered gossip networks.
\newblock {\em IEEE Jour. on Selected Areas in Information Theory}, 3(1):85--97, March 2022.

\bibitem{gossip_grid}
A.~Srivastava and S.~Ulukus.
\newblock Age of gossip on a grid.
\newblock In {\em Allerton Conference}, September 2023.

\bibitem{age_bipartite}
T.~J. Maranzatto.
\newblock Age of gossip in random and bipartite networks.
\newblock Available at arXiv2401.11580.

\bibitem{age_percolation}
T.~J. Maranzatto and M.~Michelen.
\newblock Age of gossip from connective properties via first passage percolation.
\newblock Available at arXiv:2409.12710.

\bibitem{Kaswan_mutation}
P.~Kaswan and S.~Ulukus.
\newblock Information mutation and spread of misinformation in timely gossip networks.
\newblock In {\em IEEE Globecom}, December 2023.

\bibitem{auffinger2017fifty}
A.~Auffinger, M.~Damron, and J.~Hanson.
\newblock {\em 50 years of first-passage percolation}, volume~68 of {\em University Lecture Series}.
\newblock American Mathematical Society, Providence, RI, 2017.

\bibitem{BHAMIDI_2011}
S.~Bhamidi, R.~Van Der~Hofstad, and G.~Hooghiemstra.
\newblock First passage percolation on the {Erdős–Rényi} random graph.
\newblock {\em Combinatorics, Probability and Computing}, 20(5):683–707, June 2011.

\end{thebibliography}

\if{false}
\begin{defn}
A network $G$ is $p$\textit{-good} if $P(G) \geq \frac{2}{3}$.
\end{defn}
The exact choice of $2/3$ is not important, since for any fixed value of $p$ the Markov chain $M$ mixes exponentially quickly to the absorbing state $\cF$.  We are now ready to state the main result of this section.

\begin{thm}
    If $p > $ then $C_n$/$K_n$ is $p(n)$-good.
\end{thm}
\fi
\end{document}